\documentstyle[preprint,aps]{revtex}
\begin{document}
\title{Jack polynomials, generalized binomial coefficients and polynomial 
solutions of the generalized Laplace's equation}
\author{S. Chaturvedi \thanks{email~:~~scsp@uohyd.ernet.in}}
\address{School of Physics\\
University of Hyderabad\\
Hyderabad - 500 046 (INDIA)}
\maketitle
\begin{abstract}
We discuss the symmetric homogeneous polynomial solutions of the generalized 
Laplace's equation which arises in the context of the Calogero-Sutherland 
model on a line. The solutions are expressed as linear combinations of 
Jack polynomials and the constraints on the coefficients of expansion 
are derived.  These constraints involve generalized binomial coefficients 
defined through Jack polynomials. Generalized binomial coefficients 
for partitions of $k$ upto $k=6$ are tabulated. 
\end{abstract}
\vskip1.5cm
\noindent 
\newpage
In recent years, a considerable progress has been made in
finding the exact eigen functions of Calogero-Sutherland type
models [1-6] both through operator  as well as analytical
methods [7-15] . This has largely become possible through a
better understanding of the mathematical properties of the Jack
polynomials $J_{\kappa}(x_1,\cdots, x_N;\alpha)$ [15-18], which
are homogeneous symmetric polynomials of degree $k$ in the $N$
variables $x \equiv x_1, \cdots, x_N$.  They are labelled by
partitions $\kappa$ of $k$ and depend on a parameter $\alpha$
and, among other things, are eigenfunctions of the operator
\begin{equation}
D_2 =\sum_{i}  x_{i}^2 \frac{{\partial}^2}{\partial x_{i}^{2}} 
+\frac{2}{\alpha}
\sum_{i< j} \frac{1}{(x_i-x_j)}\left( x_{i}^2\frac{\partial}{\partial x_i}
-x_{j}^2\frac{\partial}{\partial x_j}\right)~~~~,
\end{equation}
corresponding to the eigenvalue 
\begin{equation}
 E_\lambda = \sum_{i} \left[\kappa_i (\kappa_i-1) -\frac{2}{\alpha} (i-1))
\right] 
+\frac{2}{\alpha} k(N-1)~~~~,
\end{equation}
and of the Euler operator 
\begin{equation}
E_1 = \sum_i x_i \frac{\partial}{\partial x_i}~~~,
\end{equation}
corresponding to the eigenvalue $k$. 
By repeatedly calculating the commutator of the operator $D_2$ with the 
operator 
\begin{equation}
E_0 = \sum_{i} \frac{\partial}{\partial x_i}~~~,
\end{equation}
one can construct two other useful operators 
\begin{equation}
D_1 =\frac{1}{2} [ E_0, D_2] 
=\sum_{i}  x_{i} \frac{{\partial}^2}{\partial x_{i}^{2}} 
+\frac{2}{\alpha}
\sum_{i< j} \frac{1}{(x_i-x_j)}\left( x_{i}\frac{\partial}{\partial x_i}
-x_{j}\frac{\partial}{\partial x_j}\right)~~~~,
\end{equation}
\begin{equation}
D_0 =[ E_0, D_1] 
=\sum_{i}   \frac{{\partial}^2}{\partial x_{i}^{2}} 
+\frac{2}{\alpha}
\sum_{i< j} \frac{1}{(x_i-x_j)}\left( \frac{\partial}{\partial x_i}
-\frac{\partial}{\partial x_j}\right)~~~~.
\end{equation}
The operator $D_0$ is referred to as the generalized Laplacian operator.
In order to compute the action of $E_0$ on the Jack polynomials and hence 
of $D_1$ and $D_0$, it proves convenient to introduce the notion of 
generalised binomial coefficients [19] associated with partitions as follows
\begin{equation}
{\cal J}_{\kappa}(x+1)=\sum_{s=o}^{k}\sum_{\sigma} 
{\left( \begin{array}{c} \kappa \\ \sigma 
\end{array} \right)}_{\alpha} {\cal J}_{\sigma}(x)~~~,
\end{equation}
where
\begin{equation}
{\cal J}_{\kappa}(x) = \frac{J_{\kappa}(x)}{J_{\kappa}(1)}~~~,
\end{equation}
and the sum over $\sigma$ in (7) denotes sum over all partitions of $s$.
The action of the operator $E_0$ on Jack polynomials can now be written as 
\begin{equation}
E_0 {\cal J}_{\kappa}(x+1)=\sum_{i}^{N} 
{\left( \begin{array}{c} \kappa \\ \kappa_{(i)} 
\end{array} \right)}_{\alpha} {\cal J}_{\kappa_{(i)}}(x)~~~,
\end{equation}
where $\kappa_{(i)}$ is the partition obtained by decreasing the $i^{th}$ part 
of $\kappa$ by $1$ provided the result is an admissible partition. 

By making the replacements 
\begin{equation}
x^2\frac{d^2}{dx^2}~\rightarrow~ D_2~~;~~
x\frac{d^2}{dx^2}~\rightarrow~ D_1~~;~~
\frac{d^2}{dx^2}~\rightarrow~ D_0~~;~~
x\frac{d}{dx}~\rightarrow~ E_1~~;~~
\frac{d}{dx}~\rightarrow~ E_0~~,
\end{equation}
in the appropriate second order differential equations obeyed by the 
classical polynomials, new one-parameter family of multivariate special 
functions have been defined [20] and their properties have been investigated 
in detail [13]. ( These multivariate special functions subsume those defined 
through zonal harmonics [21-23] as a special case corresponding to 
$\alpha= 2$) In this context, ${\cal J}_{\kappa}(x)$ plays the same role 
as $x^k$ does in the single variable case. These multivariate 
generalised special functions appear as a factor in the exact 
eigenfunctions of Calogero-Sutherland type models. Thus, it is 
now known that the eigenfunction of Calogero-Sutherland model on a line 
involving quadratic and inversely quadratic potentials can be 
expressed as a product of (a) a Jastrow factor (b) a Gaussian and 
(c) a generalised Hermite polynomial [12,13]. However, as shown by Calogero 
[1] and Perelemov [5,6], the exact eigenfunctions of this model can also be 
expressed as a product of (a) a Jastrow factor (b) a Gaussian 
(c) a Laguerre polynomial and (d) symmetric polynomial solutions  
of the generalised Laplace's equation 
\begin{equation}
D_0 P_k = 0~~~.
\end{equation}
(This equation also turns up in the conext of the Calogero-Sutherland 
model involving only the inversely quadratic potentials). Finding 
symmetric polynomial solutions of this equation is our main concern 
here. In principle, one could expand $P_k$ in terms of any basis 
in the space of symmetric polynomials in $N$ variables and work out 
the constraints on the coefficients of expansion arising from 
the requirement that $P_k$ satisfy the generalised Laplace's equation. 
In the past, monomial symmetric functions and power symmetric 
functions have been used for this purpose. However, no explicit 
expression for the constraints on the coefficients of expansion 
could be given because the generalised Laplacian operator has no 
simple action on these functions.  In this letter we suggest that 
the most appropriate basis functions for this purpose are the 
Jack polynomials ${\cal J}_{\kappa}(x)$ as can be seen from the following 
considerations.

From the known action of $D_2$ and $E_0$ on ${\cal J}_{\kappa}(x)$ one finds 
that the actions of $D_1= 1/2~[E_0,D_2]$ and $D_0= [E_0,D_1]$
on ${\cal J}_{\kappa}(x)$ are given by 
\begin{equation}
D_1 {\cal J}_{\kappa}(x) = \sum_{i}^{N} 
{\left( \begin{array}{c} \kappa \\ \kappa_{(i)} 
\end{array} \right)}_{\alpha} (\kappa_i -1+ \frac{1}{\alpha}(N-i))
{\cal J}_{\kappa_{(i)}}(x)~~~,
\end{equation}    
\begin{eqnarray}
& &D_0 {\cal J}_{\kappa}(x) = \sum_{i}^{N} 
{\left( \begin{array}{c} \kappa \\ \kappa_{(i)} 
\end{array} \right)}_{\alpha}{\left( \begin{array}{c} \kappa_{(i)}  
\\ \kappa_{(i,i)} \end{array} \right)}_{\alpha} 
{\cal J}_{\kappa_{(i,i)}}(x)  \nonumber \\ 
&+&
 \sum_{i<j}^{N} 
\left[{\left( \begin{array}{c} \kappa \\ \kappa_{(i)} 
\end{array} \right)}_{\alpha}{\left( \begin{array}{c} \kappa_{(i)} \\ 
\kappa_{(i,j)} 
\end{array} \right)}_{\alpha}-
{\left( \begin{array}{c} \kappa \\ \kappa_{(j)} 
\end{array} \right)}_{\alpha}{\left( \begin{array}{c} \kappa_{(j)} \\ 
\kappa_{(i,j)}
\end{array} \right)}_{\alpha}\right]
((\kappa_i - \kappa_j)-\frac{1}{\alpha}(i-j)) 
{\cal J}_{\kappa_{(i,j)}}(x)~.
\end{eqnarray}    
Here $\kappa_{(i,j)}$ denotes the sequence obtained by decreasing the 
$i^{th}$ and $j^{th}$ part of the partition $\kappa$ by $1$ provided 
the result of this operation is an admissible partition. Armed with the 
knowledge of the action of the generalised Laplace's operator on 
${\cal J}_{\kappa}(x)$, we now expand $P_k$ in terms of these as follows 
\begin{equation}
P_k (x) = \sum_{\kappa} a_{\kappa} {\cal J}_{\kappa}(x)~~~,
\end{equation}
wkere the sum on the rhs runs over all partitions of $k$. Substituting this 
expansion in $(11)$, using $(13)$, and equating the  coefficients 
of the ${\cal J}$'s to zero we obtain the following constraint on the 
$a_{\kappa}$'s
\begin{eqnarray}
& & \sum_{i<j}^{N} a_{\kappa^{(i,j)}}
\left[{\left( \begin{array}{c} \kappa^{(i,j)} \\ \kappa^{(j)} 
\end{array} \right)}_{\alpha}{\left( \begin{array}{c} \kappa^{(j)} \\ \kappa 
\end{array} \right)}_{\alpha}-
{\left( \begin{array}{c} \kappa^{(i,j)} \\ \kappa^{(i)} 
\end{array} \right)}_{\alpha}{\left( \begin{array}{c} \kappa^{(i)} \\ 
\kappa
\end{array} \right)}_{\alpha}\right]
((\kappa_{i}^{(i,j)} - \kappa_{j} ^{(i,j)})-\frac{1}{\alpha}(i-j))\nonumber \\
&+ & \sum_{i}^{N}a_{\kappa^{(i,i)}} 
{\left( \begin{array}{c} \kappa^{(i,i)} \\ \kappa_{(i)} 
\end{array} \right)}_{\alpha}{\left( \begin{array}{c} \kappa_{(i)}  
\\ \kappa \end{array} \right)}_{\alpha}=0~~. 
\end{eqnarray}    
Here $\kappa^{(i)}$ ($\kappa^{(i,j)}$) denote partitions obtained by 
increasing the $i^{th}$ part ($i^{th}$ and $j^{th}$ parts) of $\kappa$ by one. 
Using these constraints in $(14)$ one obtains the linearly independent 
symmetric homogeneous polynomial solutions of the generalized 
Laplace's equation.  

The method outlined above for finding the linearly independent solution 
of the generalized Laplace's equation requires the knowledge of the 
generalized binomial coefficients. Some general properties of these 
which follow from the duality and the interpolational properties  
of Jack polynomials are listed below

\noindent
(a) The duality of Jack polynomials i.e. the algebraic homomorphism 
${\cal J}_{\kappa}(x; \alpha) \rightarrow {\cal J}_{\kappa^\prime}(x;1/\alpha)
$ where $\kappa^\prime$ denotes the partition conjugate to $\kappa$,  
implies that  
\begin{equation}
{\left( \begin{array}{c} \kappa \\ \sigma 
\end{array} \right)}_{\alpha} = {\left( \begin{array}{c} \kappa^\prime \\ 
\sigma^\prime \end{array} \right)}_{1/\alpha}~~~.
\end{equation}

\noindent
(b) For $\alpha = \infty,~2~,~1~,~0~$ Jack polynomials reduce to monomial 
symmetric functions, zonal harmonics, Schur functions and the elementary 
symmetric functions
\begin{equation}
{\cal J}_{\kappa}(x; \infty)={\cal M}_{\kappa}(x)~;~
{\cal J}_{\kappa}(x; 2)={\cal C}_{\kappa}(x)~;~
{\cal J}_{\kappa}(x;1)={\cal S}_{\kappa}(x)~;~
{\cal J}_{\kappa}(x; 0)={\cal E}_{\kappa^\prime}(x)~.
\end{equation} 

The generalised binomial coefficients for the elementary symmetric 
functions are easy to calculate and one finds that 
\begin{equation} {\left( \begin{array}{c} \kappa \\ \sigma
\end{array} \right)}_{0}=\sum_{distinct~ perm.~ of~ \sigma_{1}^{\prime}, 
\cdots, 
\sigma_{N}^{\prime}}
{\left( \begin{array}{c} \kappa_{1}^{\prime} \\ \sigma_{1}^{\prime} 
\end{array} \right)}
\cdots
{\left( \begin{array}{c} \kappa_{N}^{\prime} \\ \sigma_{N}^{\prime} 
\end{array} \right)}~~~,
\end{equation}
and hence on using the duality in $(16)$
\begin{equation}
{\left( \begin{array}{c} \kappa \\ \sigma 
\end{array} \right)}_{\infty}=\sum_{distinct~ perm.~ of~ \sigma_{1}, \cdots, 
\sigma_{N}}
{\left( \begin{array}{c} \kappa_{1} \\ \sigma_{1} 
\end{array} \right)}
\cdots
{\left( \begin{array}{c} \kappa_{N} \\ \sigma_{N} 
\end{array} \right)}~~~.
\end{equation}
This appears to be the most natural extension of the notion of the binomial 
coefficients for integers to those for partitions. If we denote by 
$\{P\sigma\}$ 
the sequences obtained by distinct permutations of the entries in the 
partition $\sigma$ then it follows from $(19)$ that 
\begin{equation} 
{\left( \begin{array}{c} \kappa \\ \sigma 
\end{array} \right)}_{\infty} = 0~~~,
\end{equation}
if all elements of the set $\{P\sigma\}$ are such that the sequences obtained 
by subtracting the sequence $\kappa$ from them contain only non negative 
integers.  From $(19)$ it also follows that if $\kappa$ and $\sigma$ are both 
partitions of 
$k$ then 
\begin{equation} 
{\left( \begin{array}{c} \kappa \\ \sigma 
\end{array} \right)}_{\infty} = \delta_{\kappa,\sigma}~~~.
\end{equation}

We have explicitly computed the generalized binomial coefficients 
for partitions of $k$ upto $k= 5$ from the knowledge of the expressions 
for the Jack polynomials in terms of monomial symmetric functions 
given in ref 24. The results are summarized in the Tables $1-5$. 
These calculations indicate that 

\noindent 
(i) the properties $(20)$ and $(21)$ hold true for  generalized binomial 
coefficients for arbitrary $\alpha$.  

\noindent
(ii) the generalized binomial coefficients are rational functions 
of $\alpha$. 

\noindent
(iii) If $\kappa$ is a partition of $k$ and $\sigma$ denote the partitions of 
$s$ then 
\begin{equation}
\sum_{\sigma}{\left( \begin{array}{c} \kappa \\ \sigma 
\end{array} \right)}_{\alpha} = {\left( \begin{array}{c} k \\ s 
\end{array} \right)}~~~.
\end{equation}
Using (ii) and (iii) and the values of the generalised binomial coefficients 
for $\alpha=2$ [25,26] , and those for $\alpha=\infty$ and $\alpha =0$ 
we have computed the table for $k=6$. The results are displayed in Table 6. 
In a recent work Okounkov and Olshanski [26] have unravelled the general 
structure of the generalised binomial coefficients. 
 
The results given above can be used for constructing the linearly 
independent solutions $P_k$ of the generalized Laplaces's equation. 
As an illustration, explicit expressions for the constraints on the 
the coefficients of expansion for $ k=3,4$ are given below

\noindent
[1] $ k =3$
\begin{equation}
3\alpha a_{3} + \alpha(\alpha-1)a_{21} - 3a_{1^3} = 0~~.
\end{equation}

\noindent
[2] $k=4$
\begin{equation}
6\alpha(1+\alpha)a_{4} + (\alpha-1) a_{31} + 2\alpha (2+\alpha) a_{2^2}
-\alpha (3+\alpha) a_{21^2} =0~~~,
\end{equation}

\begin{equation}
\alpha (1+3\alpha) a_{31} - 2(1+2\alpha)a_{2^2} + \alpha(\alpha-1)a_{21^2} 
=0~~~. 
\end{equation}

To conclude, we have suggested a convenient basis for constructing the 
linearly independent solutions of the generalized Laplace's equation which 
arises in the context of Calogero-Sutherland models. The generalized 
binomial coefficients required for this purpose are constructed for 
partitions of $k$ for $k \leq 6$.

\vskip0.35cm
\noindent{\bf Acknowledgements:}

\noindent
I wish to thank Prof. V.Srinivasan, Dr.P.K.Panigrahi, Prof. A.K. Kapoor 
and Mr N. Gurappa for numerous discussions.
\newpage

\newpage
\noindent{\bf References}
\begin{enumerate}
\item F. Calogero, J. Math. Phys. {\bf 10}, 2191 (1969).
\item F. Calogero, {\bf 10}, 2197 (1969).
\item B. Sutherland, J. Math. Phys. {\bf 12}, 246 (1971); {\bf 12}, 251 (1971).
\item B. Sutherland,  Phys. Rev. A{\bf 4}, 2019 (1971); A {\bf 5}, 1372 (1972).
\item A.M. Perelomov, Theor. Math. Phys. {\bf 6}, 263 (1971).
\item A.M. Perelomov, {\it Generalized Coherent States}, (Springer, 1986). 
\item P.J. Gambardella, J. Math. Phys. {\bf 16}, 1172 (1975).
\item L. Brink, T.H. Hanson and M. Vasiliev, Phys. Lett. B{\bf 286}, 109 
(1992).
\item A.P. Polychronakos, Phys. Rev. Lett. {\bf 69}, 703 (1993).
\item H. Ujino and M. Wadati, J. Phys. Soc. Japan {\bf 65}, 2703 (1995); 
{\bf 65}, 653 (1996).
\item L. Lapointe and L. Vinet, Comm. Math. Phys. {\bf 178}, 425 (1996).
\item K. Sogo, J. Phys. Soc. Japan {\bf 65}, 3097 (1996).
\item T.H. Baker and P.J. Forrester, Nucl. Phys. B{\bf 492}, 682 (1997). 
\item N. Gurappa and P.K. Panigrahi, cond-mat/9710035, quant-ph/9710019.
\item H. Jack, Proc. Roy. Soc (Edinburgh) A {\bf 69}, 1 (1970)
\item R.P. Stanley Adv. in Math. {\bf 77}, 77 (1989).
\item I.G. Macdonald, {\it Symmetric
functions and Hall polynomials}, $2^{nd}$ edition, (Clarendon, Oxford, 1995).
\item F. Knop and S.Sahi, J.Int.Math.Res.Notices, issue 10, 473 (1996).
\item M. Lassalle, C.R. Acad. Sci. Paris, t. Series I {\bf 310}, 253 (1990)
\item M. Lassalle, C.R. Acad. Sci. Paris, t. Series I
{\bf 312}, 425 (1991); {\bf 312} 725 (1991); {\bf 313} 579, (1992).
\item C.S. Herz, Ann. of Math. {\bf 61}, 474 (1955).
\item A.T. James in 
{\it Theory and Applications of Special Functions} (R.A. Askey ed.), 
pp 497-520 (Academic Press, New York, 1975).
\item R.J. Muirhead, {\it Aspects of multivariate 
statistical theory} (John Wiley, New York, 1982).
\item S. Chaturvedi, {\it Symmetric functions in theoretical physics}. 
To appear 
in the proceedings of the workshop on special functions 
and differential equations (K. Srinivasa Rao ed).
\item K.C.S. Pillai and G.M. Jouris, Ann. Inst. Statist. Math {\bf21},
309 (1969).
\item A.M. Parkhurst and A.T. James in {\it Selected Tables in Mathematical 
Statistics}, (H.L. Harter and D.B. Owens eds) pp 199-388 (AMS, Providence, R.I 
, 1974).
\item A. Okounkov and G. Olshanski, J. Math. Res. Lett. {\bf 4}, 69 (1997).
\end{enumerate}
\newpage
\topmargin=1truein
\textwidth=6truein 
\textheight=9truein
\flushleft
\setlength{\oddsidemargin}{-.5 in}
\setlength{\evensidemargin}{.5 in}
\begin{center} 
Table 1 : Generalized binomial coefficients for k = 1  \\ 
\vspace{1.0 cm}
\samepage
\begin{tabular}{l c| c c c  c c}
& & & $\sigma$    \\   
&  k=1 & (0) & (1) \\
\hline 
&  &  &    \\
$\kappa$   &  (1)  & 1 & 1 \\
&  &  &    \\
\hline 
\end{tabular}
\end{center}
\vspace{1.5cm}
\begin{center}
Table 2 : Generalized binomial coefficients for k=2   \\
\vspace{0.3 cm}
\samepage
\begin{tabular}{l c| c c c  c c c c  c c}

&      &     &     & $\sigma$ &    \\   
&  k=1 & (0) & (1) & (2) & ($1^2$) \\
\hline 
&  &  & & &    \\
&  (2)  & 1 & 2 & 1 & 0 \\
$\kappa$ &  &  & & &   \\
& ($1^2$) & 1 & 2 & 0 & 1 \\
\hline 
\end{tabular}
\end{center}
\newpage
\begin{center}
\vspace{1.5cm}

Table 3 : Generalized binomial coefficients for k=3  \\ 
\vspace{0.3 cm}
\samepage
\begin{tabular}{l c| c c c  c c c c  c c}

&      &     &      &        & $\sigma$ &    &       & \\   
&  k=3 & (0) & (1)  &   (2)  & ($1^2$) & (3) &(2,1)) &($1^3$) \\
   \hline 
&  &  &  &  &  &   &  &  \\
   &  (3)  & 1 & 3 & 3 & 0 &  1 & 0 & 0 \\
&  &  &  &  &  &   &  &  \\
  $\kappa$ & (2,1) &  1   &  3    & $\frac{2+\alpha}{1+\alpha}$ & 
$\frac{1+2\alpha}{1+\alpha}$  & 0 & 1 & 0  \\
&  &  &  &  &  &   &  &  \\         
& ($1^3$) &  1   &  3    & 0 & 3 & 0 & 0 & 1 \\
&  &  &  &  &  &   &  &  \\
\hline 
\end{tabular}
\end{center}
\begin{center}
\vspace{0.5 cm}
Table 4a : Generalized binomial coefficients for k=4   \\ 
\vspace{0.3 cm}
\samepage
\begin{tabular}{l c| c c c  c c c c  c c c c c c c }

&     &     &      &        & $\sigma$ &     &       &  \\   
& k=4 & (0) & (1)  &   (2)  & ($1^2$)  & (3) & (2,1) &($1^3$)   \\
   \hline
&       &    &     &    &   &   &   &  \\ 
&  (4)  & 1  &  4  & 6  & 0 & 4 & 0 & 0  \\                  
&       &    &     &    &   &   &   &  \\   
& (3,1) & 1 &  4   & $\frac{5+ 3\alpha}{1+\alpha}$ & $\frac{1+3\alpha}
{1+\alpha}$& $\frac{2+2\alpha}{1+2\alpha}$& $\frac{2+6\alpha}{1+2\alpha}$ 
& 0 \\
&       &   &      &    &   &   &   &  \\   
$\kappa$& ($2^2$) &  1   &  4    &  $\frac{4+2\alpha}{1+\alpha}$ & 
$\frac{2+4\alpha}{1+\alpha}$ & 0 & 4 & 0  \\
&       &   &     &      &  &    &  &  \\   
& (2,$1^2$) & 1 & 4 & $\frac{3+\alpha}{1+\alpha}$ & $\frac{3+5\alpha}
{1+\alpha}$ & 0 &  $\frac{6+2\alpha}{2+\alpha}$ & $\frac{2+2\alpha}
{2+\alpha}$ \\
&       &   &     &      &   & &  & \\   
& ($1^4$) & 1 & 4 & 0 & 6 & 0 & 0 & 4  \\
& & & & & & &  &  \\
\hline
\end{tabular}
\end{center}
\newpage
\begin{center}
Table 4b : Generalized binomial coefficients for k=4   \\ 
\vspace{0.3 cm}
\samepage
\begin{tabular}{l c| c c c  c c c c  c c c c c c c }

&      &     &       &  $\sigma$ &           &        \\   
&  k=4 & (4) & (3,1) & ($2^2$)   & (2,$1^2$) & ($1^4$)  \\
   \hline
 &     &     &       &           &           &  \\ 
 &  (4)        & 1 & 0 & 0 & 0 & 0 \\                  
&      &     &       &           &           &  \\   
& (3,1) &   0 & 1 & 0 & 0 & 0 \\
&      &      &      &           &           &  \\   
$\kappa$& ($2^2$) & 0 & 0 & 1 & 0 & 0 \\
&      &      &      &           &           &  \\   
& (2,$1^2$) &  0 & 0 & 0 & 1 & 0 \\
&      &      &       &           &          &  \\   
& ($1^4$) & 0 & 0 & 0 & 0 & 1 \\
&      &      &       &           &          &  \\
\hline
\end{tabular}
\end{center}
\newpage
\begin{center}  
Table 5a  : Generalized binomial coefficients for k=5  \\ 
\vspace{0.3 cm}
\samepage
\begin{tabular}{l c| c c c  c c c c  c c c c c c c }

&      &     &      &        &  $\sigma$ &     &        &           \\   
&  k=5 & (0) & (1)  &   (2)  & ($1^2$)   & (3) & (2,1) & ($1^3$)  \\
   \hline 
&  (5)  & 1  & 5    & 10     & 0         & 10  & 0      & 0   \\              
&      &     &      &        &            &     &        &           \\
& (4,1) & 1  & 5    & $\frac{9+ 6\alpha}{1+\alpha}$ & $\frac{1+4\alpha}
{1+\alpha}$& $\frac{7+8\alpha}{1+2\alpha}$& $\frac{3+12\alpha}{1+2\alpha}$
  & 0   \\
&      &     &      &        &            &     &        &           \\  
&(3,2) & 1 & 5 & $\frac{8+4\alpha}{1+\alpha}$ & $\frac{2+6\alpha}{1+\alpha}$ & 
$\frac{4+ 2\alpha}{1+2\alpha}$ & $\frac{6+ 18\alpha}{1+2\alpha}$ & 0  \\
&      &     &      &        &            &     &        &           \\ 
$\kappa$ & (3,$1^2$) & 1 & 5 & $\frac{7+3\alpha}{1+\alpha}$ & $\frac{3+7\alpha}
{1+\alpha}$ 
& $\frac{3+ 2\alpha}{1+2\alpha}$ &  $\frac{12(1+3\alpha+\alpha^2)}{(1+2\alpha)
(2+\alpha)}$ & $\frac{2+3\alpha}{2+\alpha}$  \\
&      &     &      &        &            &     &        &           \\
& ($2^2$,1) & 1 & 5 & $\frac{6+2\alpha}{1+\alpha}$ & $\frac{4+8\alpha}
{1+\alpha}$ & 0 & $\frac{18+6\alpha}{2+\alpha}$ & $\frac{2+4\alpha}
{2+\alpha}$ \\
&      &     &      &        &            &     &        &           \\ 
& (2,$1^3$) & 1 & 5 & $\frac{4+\alpha}{1+\alpha}$ & $\frac{6+9\alpha}{1+\alpha}
$ & 0 & $\frac{12+3\alpha}{2+\alpha}$ & $\frac{8+7\alpha}{2+\alpha}$ \\
&      &     &      &        &            &     &        &           \\ 
& ($1^5$) & 1 & 5 & 0 & 10 & 0 & 0 & 10  \\
\hline 
\end{tabular}
\end{center}
\newpage
\begin{center}
Table 5b  : Generalized binomial coefficients for k=5 \\ 
\vspace{0.3 cm}
\samepage
\begin{tabular}{l c| c c c  c c c c  c c c c c c c }

&      &     &       & $\sigma$ &           &  \\   
&  k=5 & (4) & (3,1) & ($2^2$)  & (2,$1^2$) & ($1^4$)  \\
\hline 
&      &     &       &  &           &  \\
&  (5)  & 5 & 0 & 0 & 0 & 0 \\                  
&      &     &       &  &           &  \\  
& (4,1)  & $\frac{2+ 3\alpha}{1+3\alpha}$ & $\frac{3+ 12\alpha}{1+3\alpha}$
 & 0 & 0 & 0  \\
&      &     &       &  &           &  \\
& (3,2) & 0 & $\frac{4+ 2\alpha}{1+\alpha}$ & $\frac{1+ 3\alpha}{1+\alpha}$ 
& 0 & 0 \\
&      &     &       &  &           &  \\
$\kappa$& (3,$1^2$)  & 0 & $\frac{3+2\alpha}{1+\alpha}$ & 0 &
 $\frac{2+3\alpha}{1+\alpha}$ & 0 \\
&      &     &       &  &           &  \\
& ($2^2$,1)  & 0 & 0 & $\frac{3+\alpha}{1+\alpha}$ & $\frac{2+4\alpha}
{1+\alpha}$ & 0 \\
&      &     &       &  &           &  \\
& (2,$1^3$)  & 0 & 0 & 0 & $\frac{12+3\alpha}{3+\alpha}$ & $\frac{3+2\alpha}
{3+\alpha}$ \\
&      &     &       &  &           &  \\ 
& ($1^5$) & 0 & 0 & 0 & 0 & 5 \\
\hline 
\end{tabular}
\end{center}
\newpage 
\begin{center}
Table 5c  : Generalized binomial coefficients for k=5  \\ 
\vspace{0.3 cm}
\samepage
\begin{tabular}{l c| c c c  c c c c  c c c c c c c }

&      &     &       &  & $\sigma$          &           &           &   \\   
&  k=5 & (5) & (4,1) & (3,2)    & (3,$1^2$) & ($2^2$,1) & (2,$1^3$) &
 ($1^5$) \\
\hline 
&      &     &       &  &           &         & & \\
&  (5)  & 1 & 0 & 0 & 0 & 0 & 0 & 0 \\                  
&      &     &       &  &           &         & & \\  
& (4,1) & 0 &1 & 0 & 0 & 0 & 0 & 0  \\
&      &     &       &  &           &         & & \\
& (3,2) & 0 & 0 & 1  & 0 & 0 & 0 & 0 \\
&      &     &       &  &           &         & & \\
$\kappa$ & (3,$1^2$)  & 0 & 0 & 0 & 1 & 0 & 0 & 0 \\
&      &     &       &  &           &         & & \\
& ($2^2$,1)  & 0 & 0 & 0 & 0 & 1 & 0 & 0 \\
&      &     &       &  &           &         & & \\
& (2,$1^3$)  & 0 & 0 & 0 & 0 & 0 & 1 & 0  \\
&      &     &       &  &           &         & & \\ 
& ($1^5$) & 0 & 0 & 0 & 0 & 0 & 0 & 1 \\
\hline 
\end{tabular}
\end{center}
\newpage
\begin{center}
Table 6a  : Generalized binomial coefficients for k=6  \\ 
\vspace{0.3 cm}

\begin{tabular}{l c| c c c  c c c c  c c c c c c c }

&      &     &      &        &   &$\sigma$     &        &           \\   
&  k=6 & (0) & (1)  &   (2)  & ($1^2$)   & (3) & (2,1) & ($1^3$)  \\
   \hline 
&  (6)  & 1  & 6    & 15     & 0         & 20  & 0      & 0   \\              
&      &     &      &        &            &     &        &           \\
& (5,1) & 1  & 6    & $\frac{14+ 10\alpha}{1+\alpha}$ & $\frac{1+5\alpha}
{1+\alpha}$& $\frac{16+20\alpha}{1+2\alpha}$& $\frac{4+20\alpha}{1+2\alpha}$ 
 & 0   \\
&      &     &      &        &            &     &        &           \\  
&(4,2) & 1 & 6 & $\frac{13+7\alpha}{1+\alpha}$ & $\frac{2+8\alpha}
{1+\alpha}$ & $\frac{12+ 8\alpha}{1+2\alpha}$ & $\frac{8+ 32\alpha}
{1+2\alpha}$ & 0  \\
&      &     &      &        &            &     &        &           \\ 
& (4,$1^2$) & 1 & 6 & $\frac{12+6\alpha}{1+\alpha}$ & $\frac{3+9\alpha}
{1+\alpha}$ & $\frac{10+ 8\alpha}{1+2\alpha}$ &  $\frac{6
(3+11\alpha+4\alpha^2)}{(1+2\alpha)(2+\alpha)}$ & $\frac{2+4\alpha}
{2+\alpha}$  \\
&      &     &      &        &            &     &        &           \\
& ($3^2$) & 1 & 6 & $\frac{12+6\alpha}{1+\alpha}$ & $\frac{3+9\alpha}
{1+\alpha}$
 &$\frac{8+4\alpha}{1+2\alpha}$  & $\frac{12+36\alpha}{1+2\alpha}$ & 0 \\
&      &     &      &        &            &     &        &           \\ 
$\kappa$& (3,2,1) & 1 & 6 & $\frac{11+4\alpha}{1+\alpha}$ & $\frac{4+11\alpha}
{1+\alpha}$ & $\frac{(3+2\alpha)(2+\alpha)}{(1+\alpha)(1+2\alpha)}$ &
 $\frac{26+83\alpha+26\alpha^2}{(2+\alpha)(1+2\alpha)}$ & $\frac{(1+2\alpha)
(2+3\alpha)}{(1+\alpha)(1+2\alpha)}$ \\
&      &     &      &        &            &     &        &           \\ 
& (3,$1^3$) & 1 & 6 & $\frac{9+3\alpha}{1+\alpha}$ & $\frac{6+12\alpha}
{1+\alpha}$ & $\frac{4+2\alpha}{1+2\alpha}$& $\frac{6(4+11\alpha+3\alpha^2)}
{(2+\alpha)(1+2\alpha)}$ & $\frac{8+10\alpha}{2+\alpha}$ \\
&      &     &      &        &            &     &        &           \\ 
& ($2^3$) & 1 & 6 & $\frac{9+3\alpha}{1+\alpha}$ & $\frac{6+12\alpha}
{1+\alpha}$ & 0 & $\frac{36+12\alpha}{2+\alpha}$ & $\frac{4+8\alpha}
{2+\alpha}$ \\
&      &     &      &        &            &     &        &           \\ 
& ($2^2$,$1^2$) & 1 & 6 & $\frac{8+2\alpha}{1+\alpha}$ &
$\frac{7+13\alpha}{1+\alpha}$ & 0 &
$\frac{32+8\alpha}{2+\alpha}$ & $\frac{8+12\alpha}{2+\alpha}$ \\
&      &     &      &        &            &     &        &
\\ & (2,$1^4$) & 1 & 6 & $\frac{5+\alpha}{1+\alpha}$ &
$\frac{10+14\alpha}{1+\alpha}$ & 0 &
$\frac{20+4\alpha}{2+\alpha}$ & $\frac{20+16\alpha}{2+\alpha}$
\\ &      &     &      &        &            &     &        &
\\ & ($1^6$) & 1 & 6 & 0 & 15 & 0 & 0 & 20 \\
\hline 
\end{tabular}
\end{center}  
\newpage
\begin{center}
Table 6b  : Generalized binomial coefficients for k=6  \\
\vspace{0.3 cm}
\samepage
\begin{tabular}{l c| c c c  c c c c  c c c c c c c }

&      &     &       & $\sigma$ &           &  \\ &  k=6 & (4) &
(3,1) & ($2^2$)  & (2,$1^2$) & ($1^4$)  \\
\hline 
&      &     &       &  &           &  \\ &  (6) & 15 & 0 & 0 &
0 & 0 \\ &      &     &       &  &           &  \\ & (5,1)  &
$\frac{9+ 15\alpha}{1+3\alpha}$ & $\frac{6+
30\alpha}{1+3\alpha}$ & 0 & 0 & 0  \\ &      &     &       &  &
&  \\ & (4,2) &$\frac{2(2+
3\alpha)(1+\alpha)}{(1+2\alpha)(1+3\alpha)}$  &
$\frac{2(5+3\alpha)(1+4\alpha)}{(1+\alpha)(1+3\alpha)}$ &
$\frac{(1+ 3\alpha)(1+4\alpha)}{(1+\alpha)(1+2\alpha)}$ & 0 & 0
\\ &      &     &       &  &           &  \\ & (4,$1^2$)  &
$\frac{3+3\alpha}{1+3\alpha}$ &
$\frac{3(3+13\alpha+8\alpha^2)}{(1+\alpha)(1+3\alpha)}$ & 0 &
$\frac{3+6\alpha}{1+\alpha}$ & 0 \\ &      &     &       &  &
&  \\ & ($3^2$)  & 0 & $\frac{12+6\alpha}{1+\alpha}$ &
$\frac{3+9\alpha}{1+\alpha}$ & 0 & 0 \\ &      &     &       &
&           &  \\ $\kappa$& (3,2,1)  & 0 &
$\frac{3(3+2\alpha)(2+\alpha)}{2(1+\alpha)^2}$ &
$\frac{3(1+3\alpha+\alpha^2)}{(1+\alpha)^2}$ &
$\frac{3(2+3\alpha)(1+2\alpha)}{2(1+\alpha)^2}$ & 0 \\ &      &
&       &  &           &  \\ & (3,$1^3$) & 0 &
$\frac{6+3\alpha}{1+\alpha}$ & 0 &
$\frac{3(8+13\alpha+3\alpha^2)}{(1+\alpha)(3+\alpha)}$ &
$\frac{3+3\alpha}{3+\alpha}$ \\ &      &     &       &  &
&  \\ & ($2^3$) & 0 & 0 & $\frac{9+3\alpha}{1+\alpha}$ &
$\frac{6+12\alpha}{1+\alpha}$ & 0 \\ &      &     &       &  &
&  \\ & ($2^2$,$1^2$) & 0 & 0 &
$\frac{(3+\alpha)(4+\alpha)}{(1+\alpha)(2+\alpha)}$ &
$\frac{2(3+5\alpha)(4+\alpha)}{(1+\alpha)(3+\alpha)}$
&$\frac{2(3+2\alpha)(1+\alpha)}{(2+\alpha)(3+\alpha)}$ \\ &
&     &       &  &           &  \\ & (2,$1^4$) & 0 & 0 & 0 &
$\frac{30+6\alpha}{3+\alpha}$ &$\frac{15+9\alpha}{3+\alpha}$ \\
&      &     &       &  &           &  \\ & ($1^6$) & 0 & 0 & 0
& 0  & 15 \\
\hline 
\end{tabular}
\end{center}
\newpage
\begin{center}
Table 6c  : Generalized binomial coefficients for k=6 \\
\vspace{0.3 cm}
\samepage
\begin{tabular}{l c| c c c  c c c c  c c c c c c c }

&      &     &       &  &$\sigma$           &           &
&   \\ &  k=6 & (5) & (4,1) & (3,2)    & (3,$1^2$) & ($2^2$,1) &
(2,$1^3$) & ($1^5$)  \\
\hline 
&      &     &       &  &           &         & & \\ &  (6)  & 6
& 0 & 0 & 0 & 0 & 0 & 0 \\ &      &     &       &  &           &
& & \\ & (5,1) & $\frac{2+4\alpha}{1+4\alpha}$
&$\frac{4+20\alpha}{1+4\alpha}$ & 0 & 0 & 0 & 0 & 0  \\ &      &
&       &  &           &         & & \\ & (4,2) & 0 &
$\frac{4+4\alpha}{1+2\alpha}$ & $\frac{2+8\alpha}{1+2\alpha}$  &
0 & 0 & 0 & 0 \\ &      &     &       &  &           &         &
& \\ & (4,$1^2$)  & 0 & $\frac{6+6\alpha}{2+3\alpha}$ & 0 &
$\frac{6+12\alpha}{2+3\alpha}$ & 0 & 0 & 0 \\ &      &     &
&  &           &         & & \\ & ($3^2$)  & 0 & 0 & 6 & 0 & 0 &
0 & 0 \\ &      &     &       &  &           &         & & \\
$\kappa$ & (3,2,1)  & 0 & 0 &
$\frac{(3+2\alpha)(2+\alpha)}{2(1+\alpha)^2}$
&$\frac{(1+2\alpha)(2+\alpha)}{(1+\alpha)^2}$
&$\frac{(2+3\alpha)(1+2\alpha)}{2(1+\alpha)^2}$ & 0 & 0  \\ &
&     &       &  &           &         & & \\ & (3,$1^3$)  & 0 &
0 & 0 &$\frac{12+6\alpha}{3+2\alpha}$  & 0 &
$\frac{6+6\alpha}{3+2\alpha}$ & 0  \\ &      &     &       &  &
&         & & \\ & ($2^3$)  & 0 & 0 & 0 & 0 & 6 & 0 & 0 \\ &
&     &       &  &           &         & & \\ & ($2^2$,$1^2$)  &
0 & 0 & 0 &$\frac{8+2\alpha}{2+\alpha}$ &
$\frac{4+4\alpha}{2+\alpha}$ & 0 & 0 \\ &      &     &       &
&           &         & & \\ & (2,$1^4$)  & 0 & 0 & 0 & 0 & 0
&$\frac{20+4\alpha}{4+\alpha}$ & $\frac{4+2\alpha}{4+\alpha}$
\\ &      &     &       &  &           &         & & \\ &
($1^6$) & 0 & 0 & 0 & 0 & 0 & 0 & 6 \\
\hline 
\end{tabular}
\end{center}
\newpage 
\begin{center}
Table 6d  : Generalized binomial coefficients for k=6  \\
\vspace{0.3cm}
\samepage
\begin{tabular}{l c| c c c  c c c c  c c c c c c c }
&      &     &       & &           &         &$\sigma$         &
&         &               &          &     \\ &  k=6 & (6) &
(5,1) & (4,2)    & (4,$1^2$) & ($3^2$) & (3,2,1) & (3,$1^3$) &
($2^3$) & ($2^2$,$1^2$) & (2,$1^4$) & ($1^6$)\\
\hline 
&       &   &   &   &   &   &   &   &   &   &   &\\ &  (6)  & 1
& 0 & 0 & 0 & 0 & 0 & 0 & 0 & 0 & 0 &  0 \\ &       &   &   &   &
&   &   &   &   &   &   &\\ &  (5,1)  & 0 & 1 & 0 & 0 & 0 & 0 &
0 & 0 & 0 & 0 & 0 \\ &       &   &   &   &   &   &   &   &   &
&   &\\ &  (4,2)  & 0 & 0 & 1 & 0 & 0 & 0 & 0 & 0 & 0 & 0 & 0 \\
&       &   &   &   &   &   &   &   &   &   &   &\\ &  (4,$1^2$)
& 0 & 0 & 0 & 1 & 0 & 0 & 0 & 0 & 0 & 0 & 0 \\ &       &   &   &
&   &   &   &   &   &   &   &\\ &  ($3^2$)  & 0 & 0 & 0 & 0 & 1
& 0 & 0 & 0 & 0 & 0 & 0  \\ &       &   &   &   &   &   &   &   &
&   &   &\\ $\kappa$&  (3,2,1)  & 0 & 0 & 0 & 0 & 0 & 1 & 0 & 0
& 0 & 0 & 0  \\ &       &   &   &   &   &   &   &   &   &   &
&\\ &  (3,$1^3$)  & 0 & 0 & 0 & 0 & 0 & 0 & 1 & 0 & 0 & 0 & 0  \\
&       &   &   &   &   &   &   &   &   &   &   &\\ &  ($2^3$)
& 0 & 0 & 0 & 0 & 0 & 0 & 0 & 1 & 0 & 0 & 0 \\ &       &   &   &
&   &   &   &   &   &   &   &\\ &  ($2^2$,$1^2$)  & 0 & 0 & 0 &
0 & 0 & 0 & 0 & 0 & 1 & 0 & 0 \\ &       &   &   &   &   &   &
&   &   &   &   &\\ &  (2,$1^4$)  & 0 & 0 & 0 & 0 & 0 & 0 & 0 &
0 & 0 & 1 & 0 \\ &       &   &   &   &   &   &   &   &   &   &
&\\ &  ($1^6$)  & 0 & 0 & 0 & 0 & 0 & 0 & 0 & 0 & 0 & 0 & 1 \\
\hline 
\end{tabular}
\end{center}
\end{document}